\documentclass[12pt]{iopart}

\usepackage{graphicx}

\begin{document}

\title{Mesoscopic transport revisited}

\author{Mukunda P. Das$^1$ and Frederick Green$^2$}
\address{$^1$ Department of Theoretical Physics,
Research School of Physics and Engineering,
The Australian National University, Canberra, ACT 0200, Australia.}
\address{$^2$ School of Physics, The University of New South Wales,
Sydney, NSW 2052, Australia.}

\begin{abstract}
Having driven a large part of the decade's progress in physics,
nanoelectronics is now passing from today's realm of the extraordinary
to tomorrow's commonplace. This carries the problem of
turning proofs of concept into practical artefacts.
Better and more sharply focussed predictive modelling will be the
ultimate guide to optimising mesoscopic technology as it matures.
Securing this level of understanding needs a reassessment of the
assumptions at the base of the present state of the field. We offer a brief
overview of the underlying assumptions of mesoscopic transport.
\end{abstract}


\section{Context of mesoscopic transport}

At the mesoscopic scale and even more so below that, the main
challenge in electron-transport theory is how to characterize
the intrinsic properties of a structure when surface-to-volume ratios
are no longer negligibly small. The notion of an
``intrinsic'' property becomes ill defined
when the interfaces to the external circuitry and the
thermal surroundings compare in size with the device that is
coupled to them.

The issue is highlighted by a simple example. What, in practice, is the
real physical size of a sub-micrometre conducting channel when the
bulk scattering length in the constituent material is measured in microns?
There is no reason to take the electrically relevant ``length''
to be, say, the lithographic one. How can one tell where the device
proper ends and its interface regions begin?

Such questions must first be addressed operationally. Only after
an answer suggests itself by what is actually seen in
measurement does it make sense to seek a general picture of transport.
As well as the dynamics, a well-posed transport theory has to
subsume the topological characterization of the device
(with interfaces) through the {\em boundary conditions} that
effectively define what is an ``open system'' within quantum kinetics.

The actual physical definition of a mesoscopic device poses
a non-trivial challenge within this field, conditioning its future
development as tightly as its present one. Moreover, even deeper
issues emerge whose physical expression and action are not negotiable.
By that we mean that it remains crucial always to give a
full accounting of the global electrodynamic properties of a
conducting system, in its entirety.
The way of including these global properties is always the same
whether the device itself is classical or quantum, macroscopic or mesoscopic.

To analyse a driven conductor as a structure of finite extent is,
by that act of abstraction, to specify it as an open system.
The constraints of microscopic {\em and} global
charge conservation both apply.
The former alone, through the continuity equation,
cannot guarantee the latter when there are open boundaries
to act as effective generators and absorbers of electron flux.

Here it is the interlinked topology of driving fields and currents
that dictates the electromagnetic response.
Thus, any purely local specification of electrical quantities
(such as the chemical potential) is not enough to determine
a globally invariant quantity such as the current.
For example, let us posit that any current is to be generated by
a difference of chemical potential, set up between the leads of a
conductor. Then such a mechanism should always explicitly guarantee
the asymptotic stability of the leads {\em far from the active device}.
In other words, perfect screening (strict long-range neutrality)
must be respected.

In this paper we adopt a definite historical position:
that all of the conceptual and formal tools necessary for
meso- and nanoscopic transport already exist
within the developments of the past.
We have in mind the quantum-kinetic framework typified by
Fermi liquid theory, the Kubo formula, and the Kadanoff-Baym-Keldysh
non-equilibrium approach: not only readily accessible
but immediately applicable as the definitive benchmarks
for any and all mesoscopic calculations at the present.

In a normal, open system, finite resistance is microscopically
understandable if and only if there is a mechanism for dissipation
of the electrical energy imparted to the system. This physical
necessity is not addressed by models that admit {\em elastic}
transmission as the sole, exclusive, scattering process.

It is worthwhile to recall the sharp qualitative distinction between
tunnelling and metallic conduction. In tunnelling
\`a la Bardeen, the conductivity as a function of barrier width $d$ is

\[
\sigma \sim \exp(-2 d\sqrt{2V_0 -k^2}),
\]

\noindent
$V_0$ being the barrier height in units of $\hbar^2/m$ and
$k$ the wave number of the incoming carrier. The exponentially
decaying nature of barrier tunnelling is clear.

As commonly applied, the Landauer conductance formula
\cite{f+g,imry}
has had the widest appeal of any transport model in mesoscopics
and has been taken as the ultimate generalisation
of the earlier Bardeen picture of tunnelling transmission.
But the basic exponential form of the transmission probability
is now freely replaced with an arbitrary quantity
(so long as it does not exceed unity).
See the comments by Ferry and Goodnick
\cite{f+g}
as well as those by Frensley
\cite{frensley}.

On the other hand, metallic conduction presents a picture of transport
that is physically very different to tunnelling. It is inherently tied
to the many-body response of carriers in free, extended states and
involves {\em collective} displacements of the Fermi sea. 
For this reason, more recent attempts to import inelasticity
and dissipation post hoc into {\em one-body}
coherent-transmission models of resistance
have yet to demonstrate their conformity with conservation laws.

Dissipation depends on the strength of electron-hole pair relaxation;
that is, it is the decay of the current-current correlation function.
By contrast, some authors attempt to mimic this via decay of
the single-particle strength alone, which compromises particle conservation.
In short: for any regime where dissipation has a role,
a conserving many-body description is crucial.
With that central concept in mind, it is clear that
there already exists a completely established microscopic
context for assessing the more recent theoretical innovations.

We place specific focus in this paper on the notion of mesoscopic
electrical conductance. First, mesoscopic conductance is the key to many other
electronic phenomena at this scale and below. Second, its theoretical
basis is the one most studied with the current set of methodologies.

In the following Section we revisit the conductance theory
widely associated with Landauer and B\"uttiker. In Section 3
we examine its leading experimental confirmations with an eye
to the conceptual issues reviewed above. Of special relevance
is the interplay of theory and the two complementary experimental methods
known as ``two-terminal'' and ``four-terminal'' measurements.
Finally in Section 4 we summarize the essential lessons that can be drawn
for the present situation in mesoscopic analysis.


\section{Landauer conductance: theory and experiment}

The physical resistance of a conductor
takes its meaning only through measurement of current and voltage
directly through the contacts to the device under test.
Naturally the best contacts for such work are those that
are seen to have the smallest disturbing effects on the
inherent resistance of the sample.

For a uniform one-dimensional (1D) wire the Landauer
model interprets the conductance in terms of the transparency
of the wire in series with its contacts (see for example
\cite{f+g,imry}). 
The whole is conceived as a quantum barrier.
For a perfectly transmitting barrier
the transmission function, namely the ratio of
outgoing to incoming carrier flux, attains its maximum of
unity. In other cases there is
reflection at the source interface, and the transmission ratio
falls below its maximum.

The Landauer formula
for the ``two-terminal'' conductance (see below) is

\begin{equation}
G_2 = G_0 {\cal T} \leq G_0;~~~~ G_0 = {2e^2\over h} \approx 77.5{\rm \mu S}.
\label{e1}
\end{equation}

\noindent
The transmission function is ${\cal T}$ and $G_0$ is Landauer's quantum:
the conductance of a perfectly uniform, perfectly one-dimensional
metallic conductor constituting a perfectly transmissive
quantum barrier. Any actual scattering by the 1D barrier is represented
by ${\cal T}$.
The transmission theory of conductance is, at base, a phenomenology.
This is because the governing parameter ${\cal T}$ is never specified
microscopically, as an actual physical object within the theory.
The theory leaves to others the central task of computing the physical
answer essential to the character of the transmission function,
and to any practical computation. See, for example, the analysis by
Agra\"{\i}t {\em et al}.
\cite{agrait}.

As a practical fact ${\cal T}$ can indeed include any scattering,
whether elastic or inelastic, if it is calculated starting from
the conserving, many-body microscopics of a conductive system.
In particular the action of dissipation in mesoscopic transport
can be directly shown via the Kubo formula, from which
the characteristic Landauer conductance steps emerge
\cite{dg}.

Perhaps it is appropriate to make an historical remark here. In Landauer's
very first publication on (classical) resistivity
due to localised scattering in metallic conduction
\cite{rolf},
he included in ${\cal T}$ any 
process that could disturb the otherwise ballistic motion of
single carriers. Landauer explicitly recognised the action of
inelastic contributions to resistance; it was other writers
who went on, well after him, to claim an exclusive role for
coherence in mesoscopics.
As a result, Landauer himself felt obliged to comment on the
conceptual gap that opened up between his original conception and
its widespread reintepretations. See his opening Comment in Ref.
\cite{rolf2}
which annotates the reprint of his seminal paper, Ref.
\cite{rolf}.
Landauer notes further how few people, in the present day,
still seem to take any time to read the 1957 work.

\subsection{Two-terminal measurements}

The Landauer formula refers to the expected outcome for conductance
when the two probes of the voltmeter
span the entire structure, consisting of the device and its two
interfaces in series. In effect, the voltage probes coincide
with the pair of terminals through which the current enters
the sample and returns to the macroscopic supply circuit.
It is referred to as a {\em two-terminal} arrangement.

In this model it is always assumed that the system presents a
strictly elastic barrier to electron flow; indeed, the barrier
may be ideally transparent. In any case, dissipation cannot occur
within this description, regardless of how the conductor may be probed.

Nevertheless, all normal conductors always entail dissipation.
It follows that there must be a corresponding, steady electrical energy loss
associated with the Landauer conductance -- as with every resistive
system -- occurring at the familiar well defined rate

\begin{equation}
P = IV = G_0 {\cal T} V^2.
\label{e2}
\end{equation}

Where is $P$ actually dissipated? The power must be discharged
somewhere. Yet it cannot be released within the core of the
1D structure; nor can it be released at the interfaces,
because the Landauer theory of
two-terminal transmission excludes dissipation in the whole
barrier and in particular at ideal transmission, ${\cal T} = 1$.
In fact, perfect conduction has been achieved and widely reported
\cite{vanWees,Wharam,ciraci}.

Experiments have thus fully confirmed the existence of
a Landauer conductance. However, they provide no insights
into the concrete issue of physical dissipation.
So an apparent ``missing link'' remains between
Eqs. (\ref{e1}) and (\ref{e2}).

\subsection{Four-terminal measurements}

In principle, one may be able to eliminate
the interface series resistances entering into the two-terminal
conductance, so accessing the actual resistance of a ballistic device.
If a pair of perfect voltage probes were to be placed
across the 1D conductor at the heart of the structure, they
would directly read off the local electrostatic potentials and provide
the quantities needed to deduce the intrinsic resistance.
The additional inner probes should
be spatially separated from the current-supply contacts, which remain
asymptotically far away: a conceptual arrangement known as a
{\em four-terminal} measurement set-up
\cite{f+g,ft2}.

The essential requirement on all internal voltage probes is
``non-invasiveness''. That is: the probes must not
substantially disrupt the pattern of carrier flux within
the device before it is probed in this way. Otherwise it
would be impossible to deconvolve the
probes' influence uniquely from the raw measurements.
What makes the mesoscopic scale notoriously challenging is
precisely that interventions with probes tend strongly
to modify the internal landscape, putting any data at risk of being
too sample-specific to give systematic clues to the transport physics.

Assuming the feasibility of non-invasive probes, the second hurdle
is that one does not know {\em a priori} where to place them
because the physical distinction between the ``true'' mesoscopic
conductor and its interfaces is ill-defined.
The margin of uncertainty in locating the probes is the scale of
the screening regions associated with the Landauer dipole
\cite{footnote}.
Nevertheless, if the voltage probes
genuinely do not disrupt current flow, one can envisage making
successive tests to systematically relocate them until
measurements no longer report any voltage drops from locally changing
densities in the interface regions. One could then argue operationally
that the ``inner'' conductor had been marked out.

Granted all of those conditions, there is a Landauer
four-terminal conductance formula for expressing the true,
intrinsic conductance of a uniform one-dimensional wire. Thus

\begin{equation}
G_4 = G_0 {{\cal T}\over {1 - {\cal T}}}.
\label{e3}
\end{equation}

\noindent
The physical explanation of Eq. (\ref{e3}) is detailed elsewhere
\cite{f+g}.
The formula implicitly adjusts the Fermi levels
{\em immediately adjacent} to the left and right ``boundaries''
of the inner conductor,
by removing the potential contribution from the Landauer dipole.

B\"uttiker
\cite{butt}
originally derived a general four-terminal conductance formula
in terms of the partial transmission
probabilities ${\cal T}_{ij}$ between all pairs $i,j$
of available terminals
(current leads have $i = 1,2$, voltage leads $j = 3,4$):

\begin{equation}
G_4 \equiv G_2
{({\cal T}_{31} + {\cal T}_{32})({\cal T}_{41} + {\cal T}_{42})\over
{{\cal T}_{31}{\cal T}_{42} - {\cal T}_{41}{\cal T}_{32}}}.
\label{e3.1}
\end{equation}

\noindent
As long as non-invasiveness applies, Eqs. (\ref{e3}) and (\ref{e3.1})
are equivalent. We will return to this formula shortly.

\section{Implications}

The additional information in a four-terminal measurement
can be combined with a two-terminal measurement to give the series
contact resistance, due strictly to the interfaces. In a perfect
quantum barrier where ${\cal T} = 1$, Eq. (\ref{e3}) clearly
implies infinite conductance. The intrinsic wire has no resistance.
Thus the contribution to the ideal two-terminal conductance $G_2 = G_0$,
can come only from the interfaces.

We now discuss a few examples out of the large experimental
literature, and how we can come to understand these results
with the help of available theories. It is not inappropriate
to say that certain observed features remain unexplained,
and do need further laboratory investigations.

\subsection{Perfect conduction}

An essentially perfect four-terminal measurement was reported in
the remarkable experiment by de Picciotto {\em et al}.
\cite{depic}.
Using a very high-quality 1D sample produced by cleaved-edge
overgrowth and built-in voltage probes of low invasiveness,
they detected essentially zero resistance in their core conductor,
so that the transmission factor was essentially unity.
Their direct four-terminal findings
complement the earlier purely two-terminal data by van Wees {\em et al}.
\cite{vanWees},
demonstrating perfect Landauer quantization of $G_2$
after removal of ``parasitic'' access resistances.

De Picciotto {\em et al}. also displayed
the two-terminal conductance of their sample. A noticeable shortfall
about 7\% below $G_0$ is observed in the height of the
sub-band steps in $G_2$. According to Eq. (\ref{e1})
this implies a non-ideal transmission factor: ${\cal T} = 0.93$.
On the other hand, Eq. (\ref{e3}) and their zero-resistance data
imply ${\cal T} = 1$.

Other four-probe experiments have been performed on mesoscopic
samples:

\begin{itemize}

\item
Kvon {\em et al}.
\cite{kvon}
reported a small but finite
intrinsic four-terminal resistance $G_4^{-1}$ in a diffusive
mesoscopic wire, where the bulk elastic-scattering mean free path
is comparable to, possibly less than, the sample length
so that transmission is no longer ballistic.
The presence of a stepwise structure is
nevertheless evident in the corresponding plots of $G_2^{-1}$,
although coherent transport does not apply in their regime.
It is still possible, however,
to subsume these results within the Landauer picture.

\item
A four-terminal experiment somewhat closer to Ref.
\cite{depic}
was performed by Reilly {\em et al}.
\cite{Reilly}
in a conventionally fabricated, low-disorder wire built
on two-dimensional GaAs heterojunction material. Unlike the
zero-resistance results
\cite{depic},
the four-terminal conductance of
the gate-defined structures reported here
(down to a 1D wire of zero nominal length)
is neither infinite nor even excessively large.
Instead, one sees in the raw four-terminal conductance
a series of perfect Landauer steps that seem to follow
Eq. (\ref{e1}) rather than Eq. (\ref{e3}),
despite the radical structural difference between $G_2$ and $G_4$
within the transmission theory. Curiously, it is reported
\cite{Reilly}
that this holds true even in zero-length channels.
\end{itemize}

\subsection{Negative intrinsic resistance}

A very different light on quantum
ballistic transport has been shed by recent multi-terminal
experiments
\cite{gao,kay}.
In each case certain specific resistances, measured internally to
the device-plus-lead assembly, become negative.
What is addressed here is an ``absolute''
resistance $V/I$ at the level of Ohm's linear law;
decidedly not some differential resistance $dV/dI$
(well known to become negative in strongly
non-linear transport, as in resonant-tunnelling structures)
which -- in sharp contrast to absolute resistance --
is {\em not} constrained by the thermodynamics of
power dissipation, namely the relation $P = I^2R > 0$.
Further, if $R_4 = 1/G_4$ is negative then Eq. (\ref{e3})
also shows that ${\cal T} > 1$.

The intrinsic resistance measured by Gao {\em et al}.
\cite{gao}
is theoretically identical with $R_4$ deduced from
B\"uttiker's generic description of multi-probe
conduction, Eq. (\ref{e3.1}), and therefore also with Landauer's
equivalent version Eq. (\ref{e3}) at low invasiveness
\cite{f+g}.
According to Ref.
\cite{gao}
it is the difference of two products of partial
probabilities, ${\cal T}_{31}{\cal T}_{42} - {\cal T}_{41}{\cal T}_{32}$,
which conceivably explains their negative intrinsic resistance
\cite{butt}.
If so, however, this entails a set of puzzling contradictions
\cite{comment}:

\begin{itemize}
\item
if $R_4 < 0$ then the power dissipation in the
device proper is $P = I^2 R_4 < 0$ and the device must spontaneously
be giving up energy to the rest of the circuit.

\item
If $R_4 < 0$ and ${\cal T} > 1$ then unitarity (probability)
is not conserved, even though the Landauer formula (\ref{e3}),
equivalent to the B\"uttiker formula (\ref{e3.1}),
is predicated on unitary single-electron propagation.
\end{itemize}

The second experiment by Kaya and Eberl
\cite{kay},
examined a closely related quantity, the three-terminal resistance,
with the conclusion that this too can be negative. In this set-up only one
voltage probe is placed inside the structure, while its
companion probe remains located on the far current lead. 

The implications here are the same as above.
The multi-probe B\"uttiker model of transmission strictly precludes
the three-terminal resistance from becoming negative.
If the latter theory is correct under the appropriate mesoscopic conditions,
the data seems once more to describe effects that are not
consistent with the theory.

We may sum up the negative-resistance outcomes as follows.
Under physically reasonable conditions, such data imply the
violation of particle and energy conservation in normal ballistic
systems. Within such systems the laws of linear transport
ought to hold, internally and externally and regardless of scale.
Unless the theory of coherent multi-probe transport is wrong,
the measurements must be open to other interpretations than
the notion of ohmic resistances' being negative
(but otherwise plain and ordinary).

\subsection{The 0.7 anomaly}

Finally we comment on the ``0.7 anomaly''. In this phenomenon,
a step in the two-probe conductance (anywhere in height
between 0.5 and 0.9 $G_0$ but quite often $\approx 0.7G_0$),
briefly precedes the onset of the normal, full-sized Landauer step
as the carrier density is increased above the sub-band threshold.

Such behaviour is not predicted by the non-interacting transmission
model of two-probe conductance. Therefore it is presumed to
be an effect of many-body correlations.
The existing 0.7 literature reveals a continuing unresolved debate,
not only between proponents of the two main theoretical lines
but also among the outcomes of various reported experiments.
Thus, for instance, if one carefully reads the contributions to the
recent Special Issue of {\em Journal of Physics: Condensed Matter}
\cite{pep}
one can form the reasonable impression that the 0.7 anomaly
is shrouded in mystery -- and remains so.

In one picture of the anomaly, it is
due to the spontaneous opening of a spin-polarized energy gap,
with the sub-band of spin states at lower energy being responsible
for the 0.7 precursor step. This approach nevertheless relies heavily
on the accepted phenomenology of conductance
as dissipationless transmission, so that the issues attaching
to the central place of energy loss must also apply here.

In the competing quite different picture, a local spin-polarized
impurity in the 1D conductor generates a Kondo-like resonance at an
energy near the bottom of the 1D conduction band. This somewhat
enhances the density of states, and hence the carrier flux, just before
the Fermi level accesses the normal conduction-band threshold
(where the standard Landauer conductance is observed). There is no
reason to believe that this scenario, at least, would have a problem
being cast within a microscopic framework
such as that of the Kubo response formula.

Ultimately, of course, only firmly validated measurements will
distinguish among theoretical alternatives. In the case of the
spin-polarization picture, there should be a channel-length
dependence of the observed size of the anomaly
\cite{akis}.
This is not so in the Kondo picture, which posits a much more localised
exchange-interaction effect.
But various sets of experimental data on the length dependence,
obtained to date, can be cited in support of either possibility. 
In other words: so far, there are no firm results to show.

\section{Conclusion}

The heart and soul of a mesoscopic theory reside in
its physical credibility. Securing that credibility
requires sustained, diligent and harmonious work by
experimentalists and theorists. As we have shown above, however, careful
study of some contemporary experimental works in mesoscopics
inspires less than full confidence in their potential to clear
the way to better mesoscopic modelling.

There are enough internal and mutual conflicts in the interpretation
of some recent observations to cloud the basic physical issues
rather than clarify them.
The theoretical task is not made easier when
experimental works, appearing in the record, contradict
one another and even themselves -- not to mention
the entire spectrum of available
mesoscopic models, widely accepted or otherwise.

\section*{Acknowledgement}

MPD acknowledges the Institute of Physics, Bhubaneshwar, India where
part of this work was done.

\end{document}